\documentclass[twocolumn]{aastex63}
\usepackage{diagbox}
\usepackage{multirow}

\usepackage{slashbox}
\usepackage{natbib}
\usepackage{amssymb}
\usepackage{amsmath}
\usepackage{verbatim}
\usepackage{indentfirst}
\usepackage{xcolor}
\usepackage{graphicx}
\usepackage{xcolor, soul}
\usepackage{graphicx}

\newcommand{\kms}{$\rm km\ s^{-1}$}
\newcommand{\ergs}{$\rm erg\ s^{-1}$}
\newcommand{\OIII}{[\rm{O\,\textsc{iii}}]}
\newcommand{\mbh}{$\rm M_{BH}$}
\newcommand{\mst}{$\rm M_{*}$}
\newcommand{\dispst}{$\rm \sigma_{*}$}
\newcommand{\dispoiii}{$\sigma_{\OIII}$}
\newcommand{\lumoiii}{$\rm L_{\OIII}$}
\newcommand{\lumradio}{$\rm L_{1.4\ GHz}$}
\newcommand{\lumcon}{$\rm \lambda L_{5100}$}
\newcommand{\lumbol}{$\rm L_{bol}$}

\newcommand{\jetpower}{$\rm P_{jet}$}

\newcommand{\snu}{\affil{Department of Physics \& Astronomy, Seoul National University, Seoul 08826, Republic of Korea}}
\newcommand{\cas}{\affil{CAS Key Laboratory for Research in Galaxies and Cosmology, Department of Astronomy, University of Science and Technology of China, Hefei 230026, China}}
\newcommand{\ustc}{\affil{School of Astronomy and Space Science, University of Science and Technology of China, Hefei 230026, China}}
\newcommand{\arios}{\affil{Aryabhatta Research Institute of Observational Sciences, Nainital-263001, Uttarakhand, India}}

\newcommand{\snuarc}{\affil{SNU Astronomy Research Center, Seoul National University, Seoul 08826, Republic of Korea}}

\begin{document}
\title{Investigating the Correlation of Outflow Kinematics with Radio Activity. Gas Outflows in AGNs. VII.}

\author{Ashraf Ayubinia} \snu \cas \ustc
\author{Jong-Hak Woo} \snu \snuarc
\author{Suvendu Rakshit}\arios
\author{Donghoon Son}\snu

\email{woo@astro.snu.ac.kr}

\begin{abstract}
We explore the relationship between the ionized gas outflows and radio activity using a sample of $\sim$ 6,000 AGNs at $z < 0.4$ with the kinematical measurement based on the \OIII~line profile and the radio detection in the VLA FIRST Survey. To quantify radio activity, we divide our sample into a series of binary subclasses based on the radio properties, i.e., radio-luminous/radio-weak, AGN-dominated/star-formation-contaminated, compact/extended and radio-loud/radio-quiet. None of the binary subclasses exhibits a significant difference in the normalized \OIII~velocity dispersion at a given \OIII~luminosity once we correct for the influence of the host galaxy gravitational potential. We only detect a significant difference of \OIII\ kinematics between high and low radio-Eddington ratio (L$_{1.4\ GHz}/L_{Edd}$) AGNs. In contrast, we find a remarkable difference in ionized gas kinematics between high and low bolometric-Eddington ratio AGNs. These results suggest that accretion rate is the primary mechanism in driving ionized gas outflows, while radio activity may play a secondary role providing additional influence on gas kinematics.

\end{abstract}

\section{Introduction}\label{sec:intro}
The supermassive black holes at the center of active galactic nuclei (AGNs) are speculated to grow in mass via the accretion process and provide strong influence on their host galaxies (e.g., \citealp{Kormendy2013}; \citealp{Heckman2014}; {\citealp{Padovani2017}). The relation between mass of black holes and their host galaxy properties implies a strong coupling, which is usually interpreted as a result of AGN feedback-regulated galaxy evolution (e.g., \citealp{Hopkins2008}; \citealp{Alexander2012}; \citealp{DeGraf2015}). However, the nature of AGN-galaxy connection seems more complex than the universal synchronized co-evolution scenario (e.g., \citealp{Peng2006}; \citealp{Woo2006}, \citealp{Woo2008}; \citealp{Rosario2012}; \citealp{Park2015}; \citealp{Pensabene2020}).\

AGN-driven outflows observed out to several kpc scales (e.g., \citealp{Rupke2013}; \citealp{Husemann2014},\citealp{Husemann2016}; \citealp{Bae2017}; \citealp{Smethurst2021}) are the most widely considered process for AGN feedback where outflowing gas strongly interacts with the ambient interstellar medium (ISM). Hence, it is crucial to unveil the nature of outflows to understand AGN feedback. In this context, two different modes of AGN feedback are usually proposed; a radiative (or quasar) mode and a kinetic (or radio) mode (see review by \citealp{Alexander2012}; \citealp{Fabian2012}). The former is predominantly operating in luminous AGNs, which are highly mass-accreting and radiatively efficient. In this mode, AGN radiation couples with gas via photoionization and photoheating, pushing gas to a few to tens of kpc scales (e.g., \citealp{Schonell2019}). In contrast, the latter generally holds for low luminosity AGNs, for which radio jets carry most of the energy as inflated bubbles filled with relativistic plasma, driving shocks and turbulence into the ambient gas (e.g., \citealp{Taylor2006}; \citealp{Mahony2016}). Due to their longer confinement time scale, low power jets (\jetpower~$\lesssim~10^{43}$ \ergs) are speculated to interact more efficiently with the ISM than higher power jets (\citealp{Mukherjee2016}). Although these two modes of AGN feedback describe very different mechanisms, it is often difficult to distinguish between the two in the observations, and it is possible that both mechanisms can drive gas outflows for given AGNs if the accretion rate is high and radio activity is strong at the same time.\

The ionized gas outflows are often traced by high-ionization lines, i.e., \OIII$\lambda5007$ line. While the virial motion of the gas in the gravitational field of the host galaxy is mainly responsible for determining \OIII~kinematics in the absence of strong outflows in non-AGN galaxies (e.g., see Figure 6 of \citealp{Woo2017}), many AGNs exhibit an additional non-gravitational component in the \OIII~profile, which is interpreted as a signature of outflows (e.g., \citealp{Nesvadba2008}; \citealp{Harrison2012}; \citealp{Bae2014}; \citealp{Woo2016}; \citealp{Wang2018}; \citealp{Leung2019}). 

In the case of radio AGNs, however, there are inconsistent views regarding the main driver of such outflows. For example, based on a large sample of AGNs detected by the Sloan Digital Sky Survey (SDSS), \cite{Mullaney2013} reported that the \OIII~line width prominently depends on radio luminosity at 1.4 GHz (\lumradio). {While they found a weak dependency of \OIII~line width on \lumoiii~or Eddington ratio, after accounting for the correlation between \OIII~luminosity (\lumoiii) and \lumradio, they argued that the mechanical energy of radio jets is likely to drive ionized gas outflows.
In contrast, \cite{Woo2016} investigated a large sample of Type 2 AGNs and similarly found that the \OIII~velocity dispersion (\dispoiii) increases with \lumradio. However, the authors noticed that this trend becomes much weaker once they normalize \OIII~velocity dispersion by the stellar velocity dispersion (\dispst) in order to correct for the effect of the gravitational potential of host galaxies. Therefore, no strong direct link between ionized gas kinematics and radio emission was presented. 
Since radio luminous AGNs are predominantly hosted by massive galaxies (\citealp{Best2005}; \citealp{Mauch2007}), the enhancement of the \OIII~velocity dispersion in these radio AGNs is partly due to the large gravitational potential of their host galaxies. 
Similar trends were also reported for a large sample of Type 1 AGNs by \cite{Rakshit2018}. Recently, \cite{Ayubinia2022} explored the correlation of outflow velocity with X-ray and radio properties in a sample of 183 nearby X-ray selected AGNs, finding no preference between X-ray luminosity and radio luminosity in driving high-velocity ionized gas outflows. 

Apart from these large statistical studies, various studies of individual radio AGNs favor the scenario of jet-induced outflows (e.g., \citealp{Nesvadba2008}; \citealp{Husemann2019}). Although it is not clear whether radio jet is the main driver of gas outflows in radio AGNs, it is of importance to investigate the role of the radio emission in comparison with the accretion power, by carefully constraining the radio properties, and by comparing gas outflows with radio and optical emission.}

In this paper, we present the detailed analysis on the connection of ionized gas kinematics with radio activity by extending our previous studies (\citealp{Woo2016}; \citealp{Rakshit2018}). First, we enlarge the sample by combining the radio AGNs used by \citet{Woo2016, Rakshit2018} with the radio AGNs analyzed by \cite{Mullaney2013}, accounting for the potential difference of the line fitting method and kinematical measurements employed by these studies. Second, we present the distribution of \OIII\ velocity and velocity dispersion to compare with radio luminosity and radio-Eddington ratios, exploring the effect of radio activity on the gas kinematics. Third, we divide the sample into various radio properties and investigate the connection of radio activity with gas kinematics. 
Based on these detailed analyses, we constrain the role of radio activity and accretion rate in driving ionized gas outflows in radio AGNs. 
 
In Section \ref{sec:sample} we describe the sample construction and the properties of radio AGNs. Section \ref{sec:results} presents the main results. In Section \ref{sec:discussion} we discuss the results and in Section \ref{sec:conclusion} we provide conclusions. A cosmology with $H_{0}$ = 70 \kms~Mpc$^{-1}$, $\Omega_{m}$ = 0.3 and $\Omega_{\Lambda}$ = 0.7 is used throughout the paper.\

\section{Sample and Analysis}\label{sec:sample}

\subsection{Sample Selection}\label{sec:selection}
We construct a large sample of radio sources including Type 1 and Type 2 AGNs from previous studies of ionized gas outflows. We describe Sample A and Sample B depending on the source of selection. 

{\bf Sample A}: We select $\sim$ 23,000 Type 2 AGNs from the catalog generated by \cite{Woo2016} and $\sim 5,000$ Type 1 AGNs from \cite{Rakshit2018}, who measured the kinematics of the \OIII~emission line, i.e., \OIII~velocity shift and velocity dispersion for low redshift ($z < 0.3$) AGNs selected from the SDSS. As \citealp{Woo2016} and \citealp{Rakshit2018} illustrated in detail, the kinematics of outflows were measured using the first and second moment of the \OIII~line profile (see Section \ref{sec:o3analysis}). By cross-matching these sources with the VLA FIRST Survey (catalog version 14dec174\footnote{\url{http://sundog.stsci.edu/first/catalogs/readme.html}}) with a matching radius of 5$^{\prime\prime}$, which is the beam size of the survey and corresponds to a physical scale $\sim$ 10 kpc at redshift $z = 0.1$, we identify radio sources with the FIRST detection limit (5$\sigma$ flux limit is 1 mJy, \citealp{White1997}), obtaining 4,778 radio AGNs. For each of them we collect measurements of stellar mass and star-formation rate (SFR), if available, from the MPA–JHU\footnote{\url{https://wwwmpa.mpa-garching.mpg.de/SDSS/}.} value-added catalog (\citealp{Abazajian2009}). This process yields 4,610 radio sources, including 920 Type 1 and 3,690 Type 2 AGNs. Hereafter, we call these radio AGNs Sample A.\

{\bf Sample B}: We adopt a sample of $\sim$ 24,000 Type 1 and Type 2 AGNs at $z < 0.4$ identified in the SDSS, for which \OIII\ emission line profiles are characterized by \cite{Mullaney2013}. We follow the same matching process as done for Sample A, to secure radio flux along with stellar mass and SFR measurements. This process provides a sample of 2,921 radio AGNs. Since both Sample A and Sample B are selected from the SDSS, a large fraction of radio AGNs overlap. Thus, we remove 772 Type 1 and 726 Type 2 AGN, which are already included in Sample A, finalizing a sample of 1,423 radio AGNs. We name this sample as Sample B. 
}

\begin{figure*}
\includegraphics[width=\linewidth]{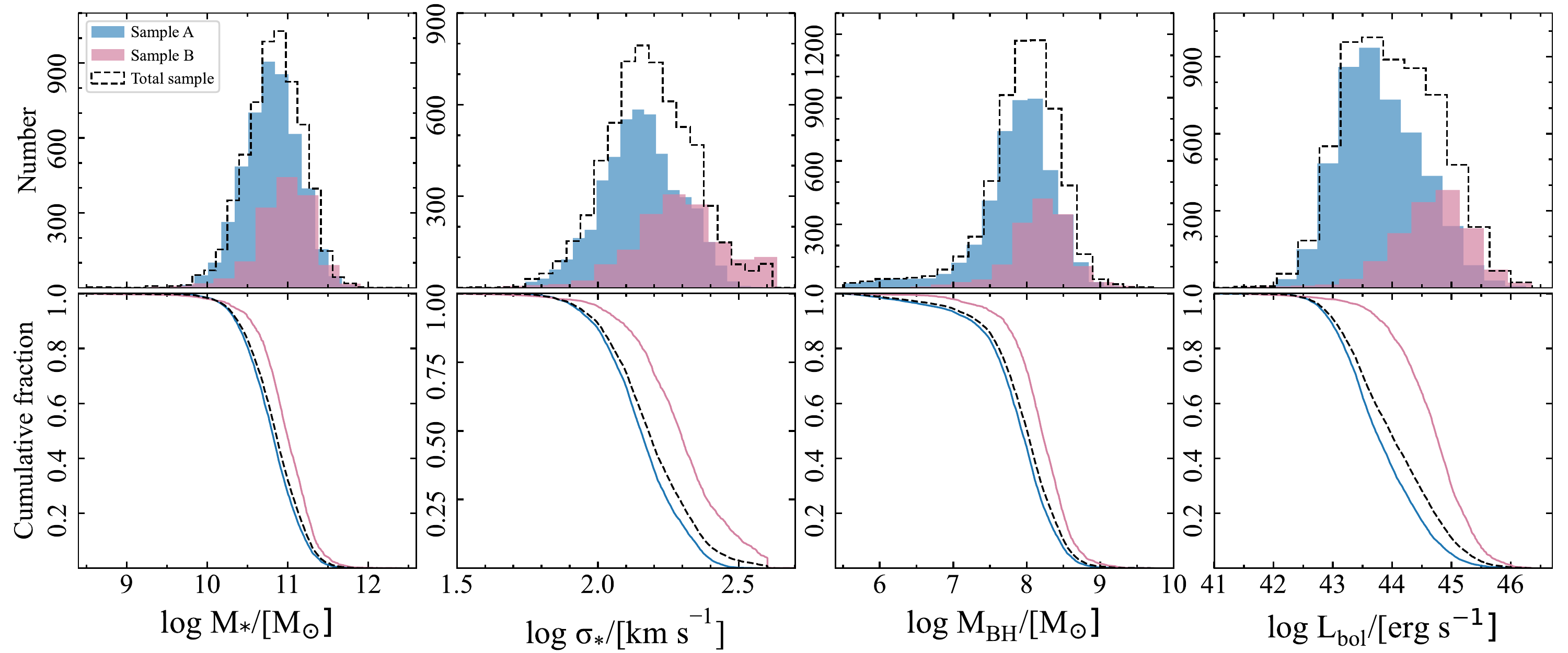}
\caption{The normalized distribution ($top$) and cumulative fraction ($bottom$) of  physical parameters for Sample A (blue), B (pink), and the combined sample (black). From $left$ to $right$: stellar mass, stellar velocity dispersion, black hole mass, and bolometric luminosity.}
\label{fig:cumulative_plots}
\end{figure*}

{\bf Total sample}: By combining Sample A and B, we construct a total sample of 6,033 radio AGNs, with well-determined \OIII~kinematics, stellar mass and SFR, among which 1,210 sources are Type 1 AGNs and 4,823 targets are Type 2 AGNs.
In Figure \ref{fig:cumulative_plots} we present the distribution of physical parameters for each sample. Note that while the estimated properties of individual objects are relatively uncertain, we use them for statistical analysis to investigate the connection of radio activity and outflow kinematics.

 %
 
\subsection{[O$_{III}$] Kinematics}\label{sec:o3analysis}
The ionized gas outflows were traced using the \OIII~emission line at 5007\AA~in both Sample A and Sample B in the previous studies. 
The details of the emission line fitting procedure were described by \citealp{Woo2016} and \citealp{Rakshit2018} (for Sample A), and \cite{Mullaney2013} (for Sample B).  Here, we briefly summarize the measurement procedure for completeness.

{\bf Sample A}: As described by \citealp{Woo2016} and \citealp{Rakshit2018}, after subtracting the host galaxy continuum (using the best-fit stellar population models) the \OIII~line profile was fitted with either a single Gaussian model or a double Gaussian model (if the peak of the second component is 3 times larger than the noise level) to measure the flux-weighted center (first moment), and velocity dispersion (second moment) of the total line profile from the best-fit model of the \OIII~line profile as
\begin{equation}\label{equ:mom1}
\lambda_{0} = {\int \lambda f_\lambda d\lambda \over \int f_\lambda d\lambda}.
\end{equation}
\begin{equation}\label{equ:mom2}
\sigma^{2}_{\rm [O\textsc{iii}]} = {\int \lambda^2 f_\lambda d\lambda \over \int f_\lambda d\lambda} - \lambda_0^2,
\end{equation}
where f$_{\lambda}$ is the flux at each wavelength. 
Since Type 1 AGNs present more complex features, i.e., AGN continuum and Fe II complex, an additional task was executed to decompose AGN continuum and FeII emission lines before the line fitting process. To model the pseudo-continuum of Type 1 AGNs, we utilized a combination of a single power law, an Fe II template, and a host galaxy template that represented for AGN continuum, Fe II complex, and host galaxy stellar contribution, respectively. We adopted the Fe II template from \citet{Kovacevic2010}, which allowed various flux ratios of the Fe II multiplet groups in the 4000-5500 \AA~spectral range \citep[for details, see][]{Rakshit2018}. 

Considering that sometimes the narrow and broad components of \OIII~profile exhibit a comparable line width, it is difficult to distinguish the wing component from the core component. Thus, we used the total \OIII~line profile to calculate the flux-weighted velocity shift (V$_{\OIII}$) and velocity dispersion ($\sigma_{\OIII}$) as a conservative approach. The velocity shift was measured with respect to the systemic velocity, which is obtained from either stellar absorption lines or the peak of the narrow component of the H$\beta$ line. Although the peak of H$\beta$ may show small additional uncertainties, it is close to the systemic velocity (see Figure 2 of \citealp{Rakshit2018}). \

To estimate the uncertainties of our measurements, we generated 100 mock spectra by randomly adding noise into the flux and obtained the best-fit for each spectrum. Then, we adopted 1$\sigma$ dispersion from mock measurements as the uncertainty. We refer the reader to \citealp{Woo2016} and \citealp{Rakshit2018} for more details.\\

{\bf Sample B}: \cite{Mullaney2013} similarly utilized either a single or double Gaussian model to fit the \OIII~emission lines. They started the fitting procedure with a single Gaussian model. Then they added another Gaussian component to improve the fit  (at a $>$ 99 per cent confidence level) for any line profile with asymmetry and/or broad component features. Since their catalog provides the measurements of the narrow and broad components individually, we utilized Equation \ref{equ:mom2} to calculate the \OIII~velocity dispersion of the total \OIII~line profile.

\subsection{Other Properties}\label{sec:main_par}
We obtain main physical properties of the sample, i.e., black hole mass (\mbh), bolometric luminosity (\lumbol), and stellar velocity dispersion (\dispst). To secure more reliable measurements, we determine these parameters for Type 1 and Type 2 AGNs separately. 

First, we estimate \mbh\ of Type 1 AGNs in Sample A based on the mass estimator given by \citet{Woo2015} using the FWHM of H$\beta$ line and the continuum luminosity at 5100 \AA~(\lumcon) as
\begin{equation}\label{equ:bh_mass1}
\rm \frac{M_{\rm BH}}{M_{\odot}} =1.12 \times 10^{6.819} (\frac{FWHM_{H\beta}}{10^{3}\ km\ s^{-1}})^{2}(\frac{\lambda L_{5100}}{10^{44}\ erg\ s^{-1}})^{0.533}.
\end{equation}
Since the measurements of the continuum luminosity at 5100 \AA~is not available for Sample B, we employ the mass estimator based on the luminosity and FWHM of H$\alpha$ (\citealp{Woo2015}):
\begin{equation}\label{equ:bh_mass2}
\rm \frac{M_{\rm BH}}{M_{\odot}} =1.12 \times 10^{6.544} (\frac{FWHM_{H\alpha}}{10^{3}\ km\ s^{-1}})^{2.06}(\frac{L_{H\alpha}}{10^{42}\ erg\ s^{-1}})^{0.46}.
\end{equation}

In the case of stellar velocity dispersion, direct measurements are difficult for Type 1 AGNs due to the strong AGN continuum. Therefore, we estimate stellar velocity dispersion from the \mbh-\dispst\ relation given by \citet{Woo2015}, which is based on the reverberation-mapped AGNs and quiescent galaxies. The bolometric luminosity for Type 1 AGNs is estimated from \lumbol~= 9 $\times$~\lumcon~(\citealp{Kaspi2000}). For Type 1 AGNs in Sample B, we estimate the continuum luminosity at 5100 \AA~from the H$\alpha$ luminosity (\citealp{Greene2005}) since the continuum luminosity is not provided by \citet{Mullaney2013}. Since less than a quarter of our Type 1 AGNs (i.e., 290 targets) is from Sample B, the additional uncertainty due to the inhomogeneous method of determining the continuum luminosity between Sample A and B makes no significant difference in the final results.\

For Type 2 AGNs, we drive the black hole mass from stellar mass (\mst) using the relation given by \citet{Marconi2003}:
\begin{equation}\label{equ:mass_sigma}
\rm \frac{M_{\rm BH}}{M_{\odot}} = 8.12 \pm 0.09 +  1.06 \pm 0.1\ (log\ \frac{M_{\rm *}}{M_{\odot}} - 10.9),
\end{equation}
We also use the direct measurements of \dispst~from the stellar absorption lines. Finally, we estimate the bolometric luminosity for Type 2 AGNs using \lumbol~= 3500 $\times$~\lumoiii, where \lumoiii~is extinction-uncorrected \OIII~luminosity. \

 %
\begin{figure*}[!htb]
\centering
\includegraphics[width=.73\linewidth]{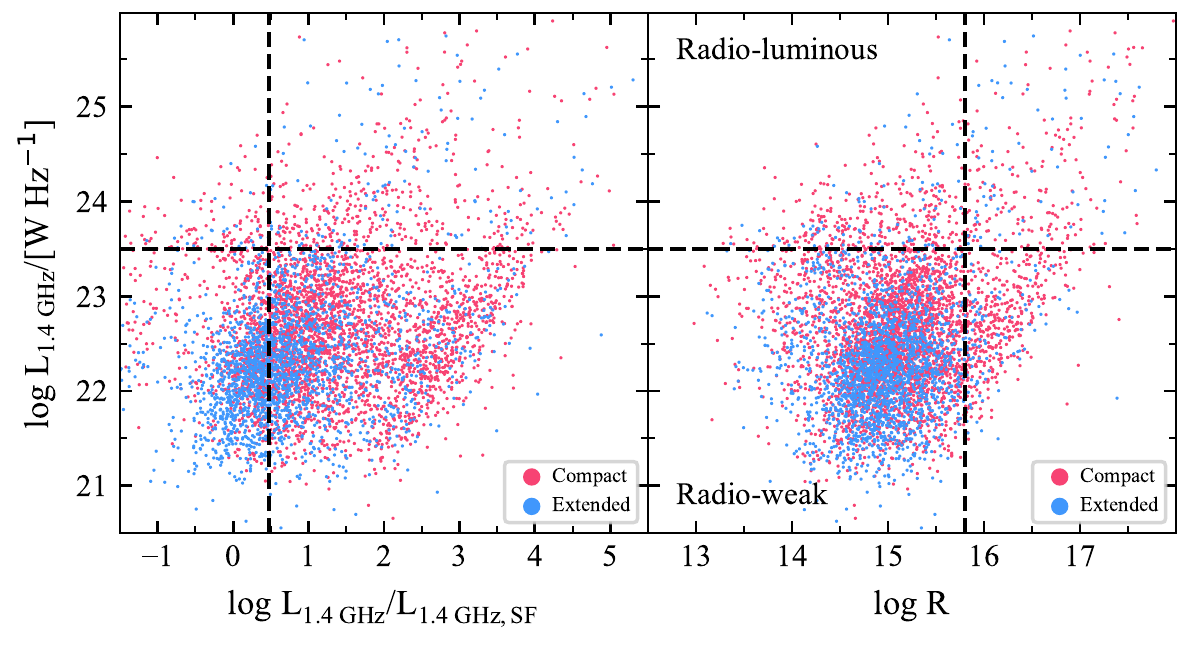}
\caption{The observed radio luminosity at 1.4 GHz (L$_{\rm 1.4\ GHz}$) as a function of the AGN dominance (the ratio of L$_{\rm 1.4\ GHz}$/L$_{\rm 1.4\ GHz,\ SF}$) ($left$) and radio loudness ($right$) for the compact (red) and extended (blue) sources. The dashed lines present the thresholds used for radio classification (see Section \ref{sec:radio_classification}).}
\label{fig:sub_class}
\end{figure*}

\subsection{Radio Classification}\label{sec:radio_classification}

We classify the radio AGNs into subclasses to investigate the connection with gas outflows. While we mainly use the radio luminosity at 1.4 GHz to trace radio activity, there is substantial contribution to the radio emission from star formation. Thus, we use four different definition of radio activity and divide the radio AGN sample into binary subclasses, respectively, based on the radio luminosity, AGN/star-formation (SF) dominance, radio compactness and radio loudness, as follows.
Note that none of these definitions perfectly distinguish strong and weak radio activities, and we will use them to find a trend with outflow kinematics.\\

{\bf 1) Radio-luminous vs. Radio-weak}:
Since the radio luminosity function below log \lumradio~$\rm \sim 23.5\ W\ Hz^{-1}$ is dominated by star-forming galaxies (SFGs) (e.g., \citealp{Best2005}; \citealp{Miller2009}; \citealp{Mauch2007}), we are motivated to divide the sample into radio-luminous and radio-weak AGNs 
at log \lumradio~$\rm = 23.5\ W\ Hz^{-1}$. We find 644 radio-luminous AGNs and 5,389 radio-weak AGNs. 
The median radio-loudness of the two subclasses is log $R \sim 15.5$ and 15.0 (see below), respectively, indicating that radio-luminous AGNs are more likely to be associated with RL AGNs. \\

{\bf 2) AGN-dominated vs. SF-contaminated}:
We distinguish AGN-dominated sources from SF-contaminated ones by comparing the observed radio luminosity with the expected radio emission from SF, in order to conservatively select pure radio AGNs. We derive the expected radio luminosity due to SF activity by using the following relation \citep{Davies2017}:
\begin{equation}\label{eq:sfr_radio}
\begin{split}
\rm log\: L_{1.4\ GHz, SF} &\rm [W\ Hz^{-1}] \sim\ 21.28 \pm 1.15\ +\\
& 1.33 \pm 0.05\ \rm log\ SFR\rm\:  [M_{\odot}\ yr^{-1}],
\end{split}
\end{equation}
where SFR is adopted from the MPA-JHU value-added catalog. These estimates were determined using the 4000 \AA~break defined as the red to blue continuum flux ratio at the 4000 \AA.
By comparing the observed radio luminosity (\lumradio) with the estimated radio luminosity from SF (L$_{\rm 1.4\ GHz,\ SF}$), we constrain the importance of AGN radio activity.  If L$_{\rm 1.4\ GHz,\ SF}$ is similar to or larger than \lumradio, then the radio emission is dominated by SF activity. In contrast, if the \lumradio~is substantially larger than the  L$_{\rm 1.4\ GHz,\ SF}$, then AGN jet is a dominant source of radio emission. Based on this assumption, we define AGN-dominated sources when \lumradio~is larger than L$_{\rm 1.4\ GHz,\ SF}$ by a factor of 3. Otherwise, we classify the source as SF-contaminated. Accordingly, we obtain 4,229 AGN-dominated and 1,804 SF-contaminated sources. Note that we choose a factor of 3 as a threshold considering the uncertainties of SFR estimates and SF-induced radio emission, to conservatively select pure radio AGNs. If we change the threshold ratio (i.e., 1 or 10) for defining subclasses, the number of targets in each subclass varies, However, the main results remain the same. 

The AGN-dominated sources have slightly higher radio luminosity and radio loudness (i.e., log \lumradio~$\sim$ 22.7 W Hz$^{-1}$ and log $R \sim$ 15.3) in comparison with SF-contaminated sources (i.e., log \lumradio~$\sim$ 22.4 W Hz$^{-1}$ and log $R \sim$ 14.6). \
\\

{\bf 3) Compact vs. Extended}:
Radio sources can be classified as compact or extended using the ratio between the integrated radio flux density (F$_{\rm int}$) and the peak radio flux density (F$_{\rm peak}$). Following \citet{Kimball2008}, we define a radio source as compact (i.e., unresolved at $\sim$ 5${''}$ resolution of FIRST) if 
\begin{equation}\label{eq:compact}
\theta = (F_{\rm int}/F_{\rm peak})^{1/2} < 1.06.
\end{equation}
Otherwise, the target is classified as an extended source. This criterion returns 3,803 (63\%) compact and 2,230 (37\%) extended radio AGNs. 
The two subclasses exhibit comparable median radio luminosities (i.e., log \lumradio~$\sim$ 22.6 vs. 22.3 W Hz$^{-1}$) as well as similar radio-loudness (i.e., log $R \sim$ 15.1 vs. 14.9, see below). Note that our main results are insensitive to the adoption of other criteria of compactness (e.g., \citealp{Yamashita2018}), while the number of sources in each class may change. 
\\

{\bf 4) Radio-Loud vs. Radio-Quiet}:
Finally, we classify radio-loud (RL) and radio-quiet (RQ) AGNs by comparing radio luminosity at 1.4 GHz with H${\alpha}$ line luminosity. While the classical definition of the RL AGNs is proposed with $R^{(K)}$ $>$ 10, where $R^{(K)}$ is the ratio of the 5 GHz radio luminosity and B-band luminosity, i.e., $R^{(K)}$ = L$\rm _{5\ GHz}/L_{B}$ (\citealp{Kellermann1989}), we use a slightly modified definition, $R$ = \lumradio~[$\rm W\ Hz^{-1}]/L_{H\alpha}\ [L_{\odot}$], where \lumradio~is the radio luminosity at 1.4 GHz in units of W Hz$^{-1}$ and $\rm L_{H\alpha}$ is H$\alpha$ luminosity in units of the solar luminosity (\citealp{Wierzbowska2017}). Assuming a radio spectral index $\alpha_{r}$ = 0.8, $\rm L_{bol}$ = 5 $\rm \nu_{B}L_{\nu_{B}}$ (\citealp{Runnoe2012}) and using Equation 1 in \citet{Wierzbowska2017}, we obtain $R^{(K)} \simeq 1.6 \times 10^{-15} R$. Therefore, RL AGNs are identified with log $R > 15.8$ and RQ AGNs with log $R < 15.8$. Based on this classification, we obtain 739 (12\%) RL and 5,294 (88\%) RQ AGNs. The median radio luminosity of the RL and RQ AGNs is log \lumradio~$\sim$ 23.0 and 22.4 W Hz$^{-1}$, respectively. The fraction of RL AGNs in our sample is consistent with that of previous studies ($\sim$ 10\%–20\%; e.g., \citealp{Jiang2007}; \citealp{Kellermann2016}; \citealp{Gupta2020}).\\

\begin{center}
\begin{table*}
\scriptsize
\caption{The fraction of AGNs (in per cent) in each subclass.}
\begin{tabular}{|c|c|c|c|c|}
\hline
\multirow{2}{*}{Radio Class} & \diagbox[innerleftsep=-10mm,innerrightsep=-0.4mm,innerwidth=5cm,dir=SW]{\textbf{Radio-luminous}}{\textbf{Radio-weak}} & 
\diagbox[innerleftsep=-10mm,innerrightsep=-0.4mm,innerwidth=5.3cm,dir=SW]{\textbf{AGN-dominated}}{\textbf{SF-contaminated}}&
\diagbox[innerleftsep=-10mm,innerrightsep=-0.4mm,innerwidth=3.5cm,dir=SW]{\textbf{Compact}}{\textbf{Extended}} & 
\diagbox[innerleftsep=-10mm,innerrightsep=-0.4mm,innerwidth=2.4cm,dir=SW]{\textbf{RL}}{\textbf{RQ}}\\
\hline
\multirow{2}{*}{\textbf{Radio-luminous}} & \diagbox[innerleftsep=-10mm,innerrightsep=-0.4mm,innerwidth=5cm,dir=SW]{100}{0} & 
\diagbox[innerleftsep=-10mm,innerrightsep=-0.4mm,innerwidth=5.3cm,dir=SW]{80}{20}&
\diagbox[innerleftsep=-10mm,innerrightsep=-0.4mm,innerwidth=3.5cm,dir=SW]{77}{23} & 
\diagbox[innerleftsep=-10mm,innerrightsep=-0.4mm,innerwidth=2.4cm,dir=SW]{39}{61}\\
\hline
\multirow{2}{*}{\textbf{Radio-weak}} & \diagbox[innerleftsep=-10mm,innerrightsep=-0.4mm,innerwidth=5cm,dir=SW]{0}{100} & 
\diagbox[innerleftsep=-10mm,innerrightsep=-0.4mm,innerwidth=5.3cm,dir=SW]{69}{31}&
\diagbox[innerleftsep=-10mm,innerrightsep=-0.4mm,innerwidth=3.5cm,dir=SW]{61}{39} & 
\diagbox[innerleftsep=-10mm,innerrightsep=-0.4mm,innerwidth=2.4cm,dir=SW]{9}{91}\\
\hline
\multirow{2}{*}{\textbf{AGN-dominated}} & \diagbox[innerleftsep=-10mm,innerrightsep=-0.4mm,innerwidth=5cm,dir=SW]{12}{88} & 
\diagbox[innerleftsep=-10mm,innerrightsep=-0.4mm,innerwidth=5.3cm,dir=SW]{100}{0}&
\diagbox[innerleftsep=-10mm,innerrightsep=-0.4mm,innerwidth=3.5cm,dir=SW]{70}{30} & 
\diagbox[innerleftsep=-10mm,innerrightsep=-0.4mm,innerwidth=2.4cm,dir=SW]{17}{83}\\
\hline
\multirow{2}{*}{\textbf{SF-contaminated}} & \diagbox[innerleftsep=-10mm,innerrightsep=-0.4mm,innerwidth=5cm,dir=SW]{7}{93} & 
\diagbox[innerleftsep=-10mm,innerrightsep=-0.4mm,innerwidth=5.3cm,dir=SW]{0}{100}&
\diagbox[innerleftsep=-10mm,innerrightsep=-0.4mm,innerwidth=3.5cm,dir=SW]{46}{54} & 
\diagbox[innerleftsep=-10mm,innerrightsep=-0.4mm,innerwidth=2.4cm,dir=SW]{1}{99}\\
\hline
\multirow{2}{*}{\textbf{Compact}} & \diagbox[innerleftsep=-10mm,innerrightsep=-0.4mm,innerwidth=5cm,dir=SW]{13}{87} & 
\diagbox[innerleftsep=-10mm,innerrightsep=-0.4mm,innerwidth=5.3cm,dir=SW]{78}{22}&
\diagbox[innerleftsep=-10mm,innerrightsep=-0.4mm,innerwidth=3.5cm,dir=SW]{100}{0} & 
\diagbox[innerleftsep=-10mm,innerrightsep=-0.4mm,innerwidth=2.4cm,dir=SW]{15}{85}\\
\hline
\multirow{2}{*}{\textbf{Extended}} & \diagbox[innerleftsep=-10mm,innerrightsep=-0.4mm,innerwidth=5cm,dir=SW]{7}{93} & 
\diagbox[innerleftsep=-10mm,innerrightsep=-0.4mm,innerwidth=5.3cm,dir=SW]{56}{44}&
\diagbox[innerleftsep=-10mm,innerrightsep=-0.4mm,innerwidth=3.5cm,dir=SW]{0}{100} & 
\diagbox[innerleftsep=-10mm,innerrightsep=-0.4mm,innerwidth=2.4cm,dir=SW]{8}{92}\\
\hline
\multirow{2}{*}{\textbf{RL}} & \diagbox[innerleftsep=-10mm,innerrightsep=-0.4mm,innerwidth=5cm,dir=SW]{34}{66} & 
\diagbox[innerleftsep=-10mm,innerrightsep=-0.4mm,innerwidth=5.3cm,dir=SW]{99}{1}&
\diagbox[innerleftsep=-10mm,innerrightsep=-0.4mm,innerwidth=3.5cm,dir=SW]{77}{23} & 
\diagbox[innerleftsep=-10mm,innerrightsep=-0.4mm,innerwidth=2.4cm,dir=SW]{100}{0}\\
\hline
\multirow{2}{*}{\textbf{RQ}} & \diagbox[innerleftsep=-10mm,innerrightsep=-0.4mm,innerwidth=5cm,dir=SW]{8}{92} & 
\diagbox[innerleftsep=-10mm,innerrightsep=-0.4mm,innerwidth=5.3cm,dir=SW]{66}{34}&
\diagbox[innerleftsep=-10mm,innerrightsep=-0.4mm,innerwidth=3.5cm,dir=SW]{61}{39} & 
\diagbox[innerleftsep=-10mm,innerrightsep=-0.4mm,innerwidth=2.4cm,dir=SW]{0}{100}\\
\hline
\end{tabular}
\label{tab:fractions}
\end{table*}
\end{center}

We summarize and compare the distributions of the radio properties based on the aforementioned criteria in Figure \ref{fig:sub_class} and Table \ref{tab:fractions}.
We find that most of radio-luminous sources (80\%) are AGN-dominated while most of SF-contaminated sources (93\%) are radio-weak. In the case of compactness, we find 77\% of radio-luminous sources are compact while 61\% of radio-weak sources are also compact, suggesting that compact-extended subclasses may not efficiently distinguish radio activity since the FIRST resolution of 5\arcsec\ is too low to identify small scale radio jets. 
We find that almost all RL AGNs ($\sim$ 99\%) are AGN-dominated, while only 17\% of AGN-dominated sources are RL AGNs.

%

\section{Results}\label{sec:results}

Ionized gas outflows are often investigated with the line profile of AGN emission lines. For example, the velocity offset of gas with respect to systemic velocity of host galaxies is measured using the peak velocity or the flux-weighted velocity (first moment of a line profile as defined by Equation \ref{equ:mom1}), and the width of emission lines, i.e., FWHM or flux-weighted velocity dispersion (second moment of a line profile as defined by Equation \ref{equ:mom2}) are used to quantify outflow kinematics. Unlike the integral field spectroscopy with a spatial resolution to measure gas kinematics as a function of the radial distance from the center, the flux-weighted spectra (i.e., SDSS spectra) are limited to providing averaged velocity to the line-of-sight or velocity dispersion. Nevertheless, such kinematical information was used to characterize the gas outflows of a large sample of AGNs.
In this section we present the ionized gas kinematics and their connection with radio luminosity and radio-Eddington radio using the flux-weighted \OIII\ velocity and velocity dispersion measurements of our radio AGN sample. 


\subsection{Outflow Kinematics and the Effect of the Gravitational Potential}\label{sec:gravity}

\begin{figure*}[!htb]
\centering
\includegraphics[width=\linewidth]{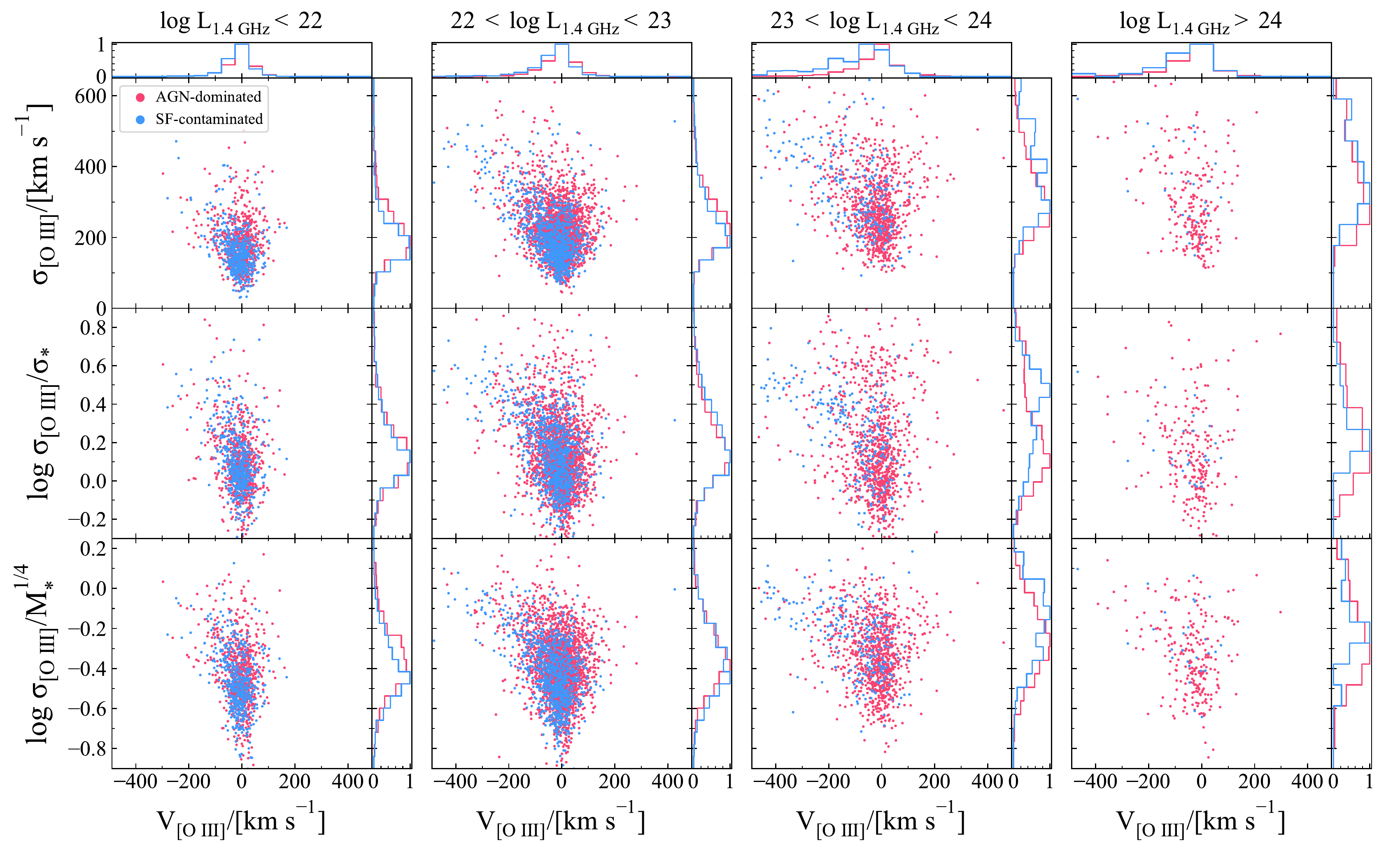}
\caption{\OIII\ velocity-velocity dispersion diagram, with and without normalization by stellar velocity dispersion and stellar mass, as a function of radio luminosity. AGN-dominated and SF-contaminated AGNs are denoted with red and blue dots, respectively.}
\label{fig:vvd1}
\end{figure*}
 \begin{figure*}[!htb]
\centering
\includegraphics[width=\linewidth]{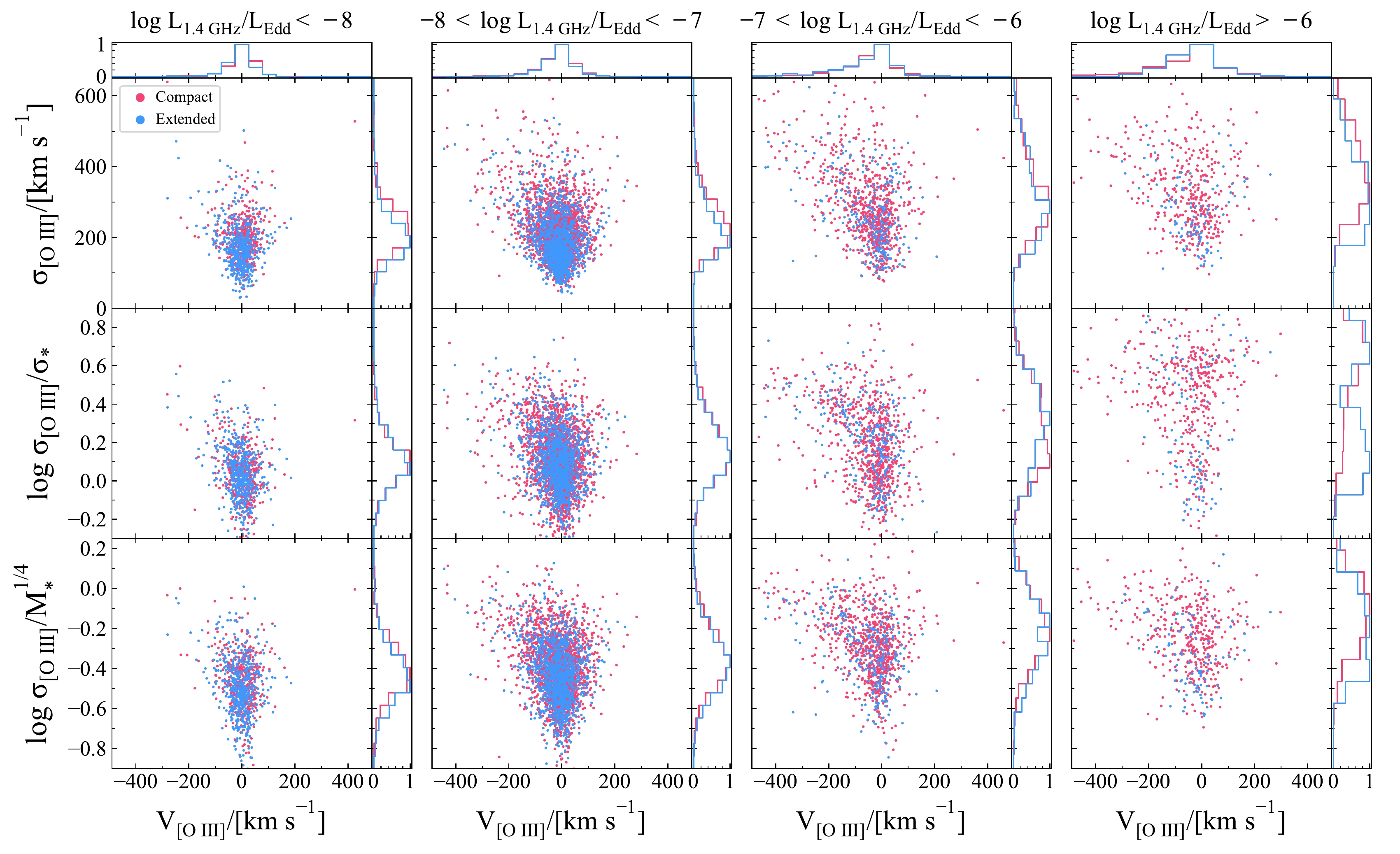}
\caption{\OIII\ velocity-velocity dispersion diagram, with and without normalization by stellar velocity dispersion and stellar mass, as a function of radio-Eddington ratio. Compact and extended sources are represented by red and blue dots, respectively.}
\label{fig:vvd2}
\end{figure*}

It has been previously demonstrated that when the \OIII~line is very broad or shows an asymmetric line profile, it can not be entirely explained by the gravitational potential of the host galaxy. For instance, when \OIII~line profiles of Type 2 AGNs exhibit an additional broad component, the gas velocity dispersion measured from the total \OIII~line profile is larger than the stellar velocity dispersion, indicating the presence of a non-gravitational component. While the stellar velocity dispersion of typical non-AGN galaxies ranges from $\sim$ 100 to 200 \kms, \OIII~velocity dispersion goes up to 600-700 \kms\ \citep{Woo2016}. On the other hand, the Doppler broadening due to the gravitational potential can be significant in massive galaxies even in the case of relatively weak or no gas outflows. Thus, it is crucial to investigate the effect of the gravitational potential and correct for in order to properly determine outflow strength. For example, \OIII~velocity dispersion can be normalized by either stellar velocity dispersion or stellar mass (e.g., \citealp{Woo2016}; \citealp{Rakshit2018}), to investigate the relative strength of outflows compared to gravitational potential. In particular, the effect of the gravitational potential can be substantially large in radio luminous AGNs because they are preferentially hosted by massive galaxies (i.e., \dispst~$\sim$ 200-300 \kms), and \OIII~lines can be relatively broad even without the presence of AGN outflows.\

In Figure \ref{fig:vvd1} we present the \OIII~velocity-velocity dispersion diagram of our radio AGNs, which are divided into four different radio luminosity bins. 
In each radio luminosity bin, we separate AGN-dominated and SF-contaminated targets as defined in Section \ref{sec:radio_classification}.
We clearly notice an increasing trend of velocity shift and velocity dispersion with radio luminosity, suggesting a connection of outflow kinematics with radio luminosity. 
When we normalize \OIII~velocity dispersion with stellar velocity dispersion or stellar mass to correct for the effect of host galaxy gravitational potential, the increasing trend with radio luminosity is somewhat suppressed. For SF-contaminated AGNs, which are mainly located in the low luminosity bins (i.e., log \lumradio~ $\rm <\ 23\ W\ Hz^{-1}$), \OIII~velocity shift is typically small and \OIII~velocity dispersion is comparable to stellar velocity dispersion, indicating no significant outflows. In contrast, the distribution of \OIII\ velocity-velocity dispersion of AGN-dominated sources is much broader, and for some objects \OIII\ velocity dispersion is clearly larger than stellar velocity dispersion.
However, this fraction is relatively small and we find no strong dependency of the \OIII\ kinematics on the radio luminosity once we correct for the effect of host galaxy gravitational potential.

We investigate the effect of the radio-Eddington ratio, which is defined as $\rm L_{1.4\ GHz}/L_{Edd}$, bottom, on the \OIII~velocity-velocity dispersion diagram in Figure \ref{fig:vvd2}. We find a clear increasing trend of velocity shift and velocity dispersion with the radio-Eddington ratios. The increasing trend does not disappear even when we normalize \OIII\ velocity dispersion with stellar velocity dispersion or stellar mass. Most of the strong outflows (i.e., $\sigma_{OIII}$/$\sigma_*$ $>$ 3) are found in high radio-Eddington ratio bins, and the velocity shift also shows a clear increasing trend with increasing radio-Eddington ratios. 
In each bin we separate compact and extended sources as defined in Section \ref{sec:radio_classification}. Note that most of extended radio sources are located in low radio-Eddington bins  with no strong outflows. In contrast, compact sources are dominant among high radio-Eddington ratio AGNs. 

Based on these results, we conclude that while the previously reported correlation between \OIII\ line width (i.e., velocity dispersion) and radio luminosity is partly due to the effect of host galaxy gravitational potential, radio activity plays a role in driving gas outflows as \OIII velocity shift and velocity dispersion increases with increasing radio luminosity with respect to the Eddington luminosity.

\begin{figure}[!htb]
\includegraphics[width=\linewidth]{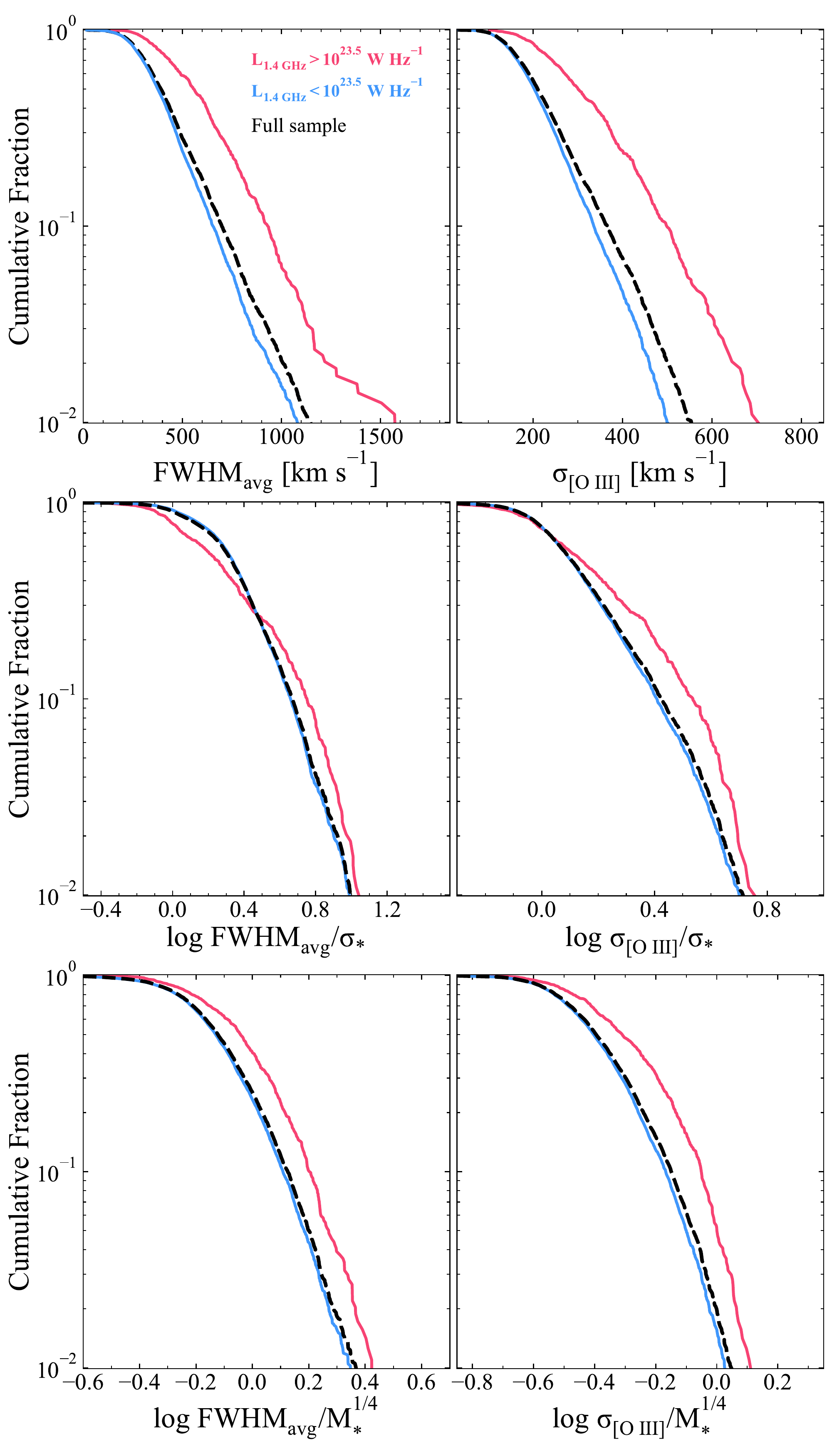}
\caption{The fractional distribution of non-normalized \OIII~line width ($top$) and \OIII~line width normalized by either stellar velocity dispersion ($middle$) or stellar mass ($bottom$) for radio-luminous (red) and radio-weak (blue) AGNs as well as our full sample (black). $Left$: flux weighted full-width-at-hafl-maximum (FWHM) and $right$: velocity dispersion.}
\label{fig:cum}
\end{figure}

To further investigate the effect of host galaxy gravitational potential on the \OIII\ line profile, we investigate the cumulative fraction of the \OIII~line width, respectively for radio-luminous and radio-weak AGNs. First, we estimate flux weighted full-width-at-hafl-maximum ($\rm FWHM_{avg}$) as used by \citet{Mullaney2013} 

\begin{equation}\label{eq:fwhm_avg}
\begin{split}
\rm FWHM_{avg} = \sqrt{(FWHM_{A}F_{A})^{2} + (FWHM_{B}F_{B})^{2}}
\end{split}
\end{equation}
where $\rm F_{A}$ and $\rm F_{B}$ are the fractional fluxes of the broad and narrow components. In top panels of Figure \ref{fig:cum} we find that the occurrence of AGNs with a broad \OIII~profile ($\rm FWHM_{avg} > 1,000$ \kms~or $\rm \sigma_{\OIII} > 400$ \kms ) is $\sim 4$ times more frequent in radio-luminous AGNs compared to either radio-weak AGNs or the total sample. This finding is in agreement with \cite{Mullaney2013} who reported that the fraction of AGNs with \OIII~$\rm FWHM_{avg} >$ 1,000 \kms~in radio-luminous AGNs is higher by a factor of $\sim$ 5 than in their total sample (including radio-undetected AGNs) based on the same practice of dividing radio-luminous and radio weak AGNs. 
However, this striking contrast is substantially weakened once we normalize \OIII\ line width by either stellar velocity dispersion (middle) or stellar mass (bottom). For example, we find that the fraction of strong outflows (i.e, $\rm log\ \sigma_{\OIII}/\sigma_{*} > 0.3$ or comparabley, $\rm log\ \sigma_{\OIII}/M_{*}^{1/4} > -0.3$) in radio-luminous AGNs is higher than radio-weak AGNs or the total sample by only a factor of $\sim 1.5$. Note that if we instead use the flux-weighted average ($\rm FWHM_{avg}$) of two FWHMs from double Gaussian models as adopted by \cite{Mullaney2013}, the result remains the same. We also note that the same conclusion is given if we only use the sample from \cite{Mullaney2013}. These results support that radio activity is playing a role in driving ionized gas outflows although it may not be a dominant role.

 %

\subsection{Ionized Gas Outflows vs. Radio Emission}\label{sec:radio}

In this Section, we investigate the connection of radio activity with outflow kinematics by comparing \OIII\ velocity dispersion between radio subclasses that we define
in Section \ref{sec:radio_classification}. Note that since none of the four radio classifications clearly separate AGN radio activity from contamination, we use each radio classification to explore a trend with gas outflows.

\begin{figure}[!htb]
\centering
\includegraphics[width=0.95\linewidth]{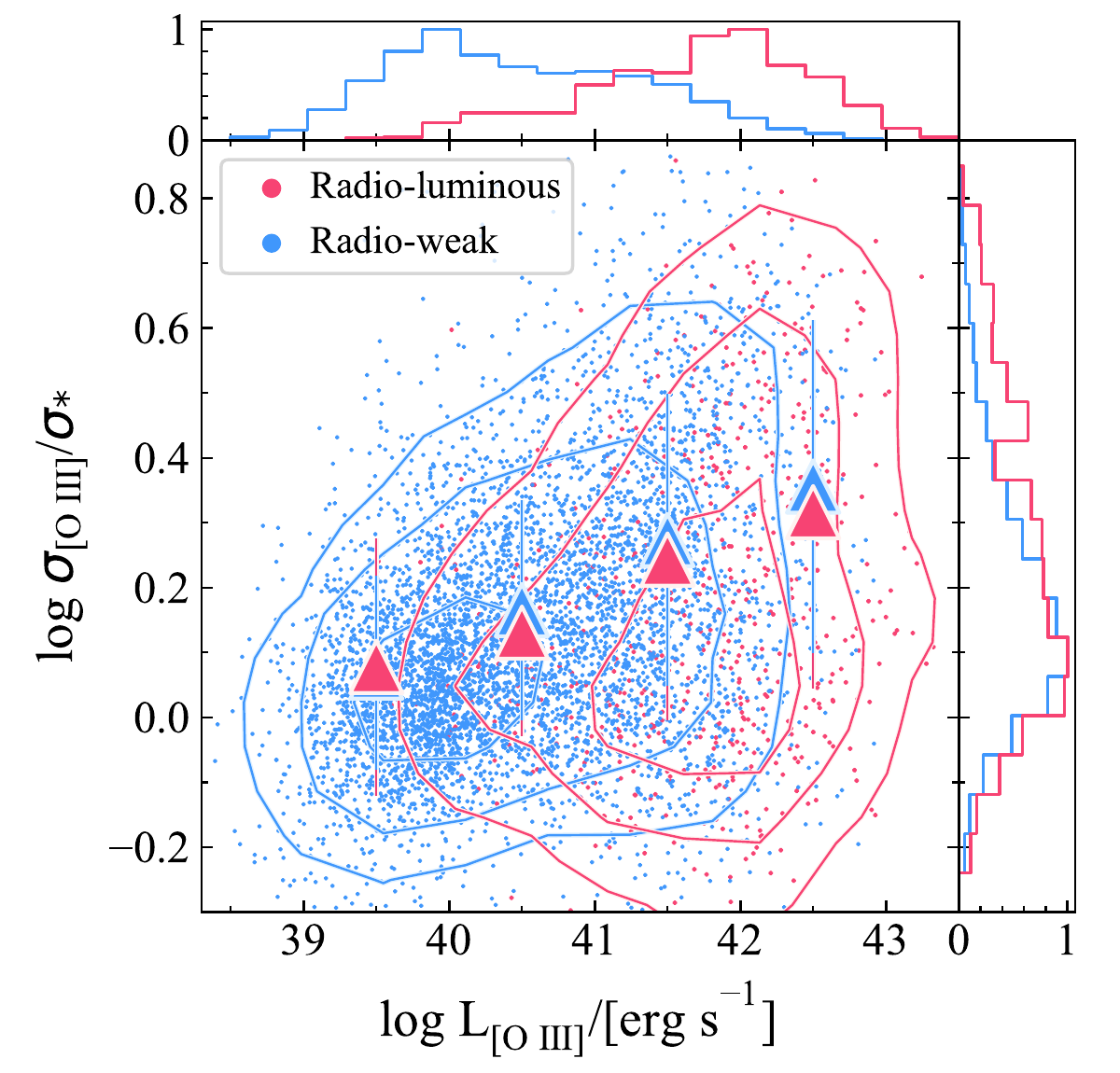}
\caption{The normalized \OIII~velocity dispersion as a function of \OIII~luminosity for radio-luminous (red) and radio-weak AGNs (blue). The mean value of \dispoiii/\dispst~in each \lumoiii~bin (with $\rm \Delta$log \lumoiii~= 1 dex) is denoted by triangles. The contours enclose 30\%, 70\% and 90\% of targets in each subsample. The error bars indicate 1$\sigma$ dispersion.}
\label{fig:luminosity}
\end{figure}

First, we compare radio-luminous and radio-weak AGNs to investigate whether radio luminosity is directly connected with outflow kinematics. In Figure \ref{fig:luminosity}  we present the normalized \OIII\ velocity dispersion as a function of \OIII\ luminosity. While we find an increasing trend of \OIII\ velocity dispersion with increasing \OIII\ luminosity in the total sample, we find no significant difference of this trend between radio-luminous and radio-weak subsamples. While radio-luminous subsample has higher mean \OIII\ luminosity than radio-weak subsample, the normalized \OIII\ velocity dispersion of the two subsamples are almost same at fixed \OIII\ luminosity, indicating that radio luminosity is not directly connected to outflow kinematics. 

 %

\begin{figure}[!htb]
\centering
\includegraphics[width=0.95\linewidth]{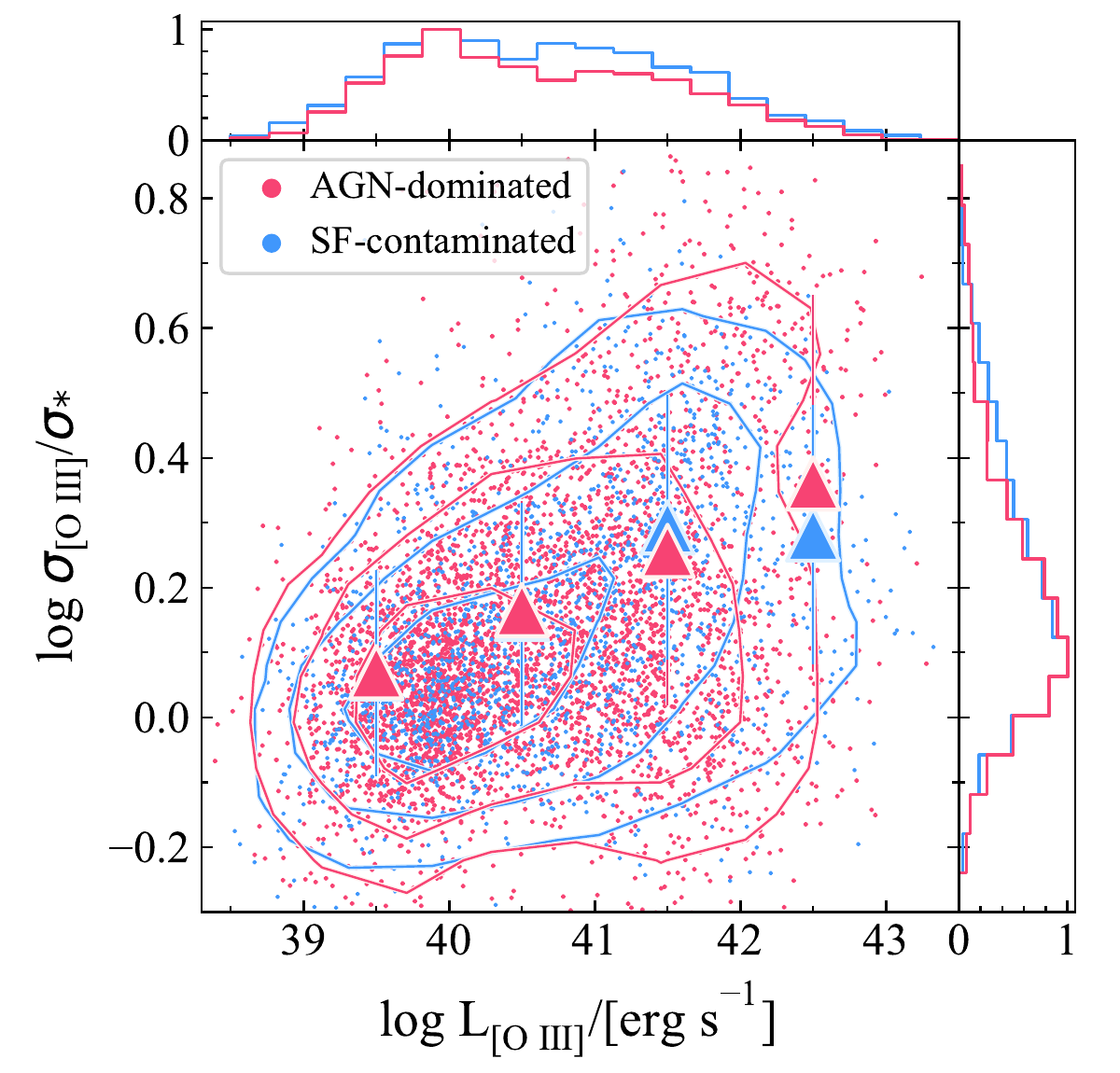}
\caption{Same as in Figure \ref{fig:luminosity}, but for AGN-dominated (red) and SF-contaminated AGNs (blue).}
\label{fig:AGN_dominated}
\end{figure}

Second, we compare  AGN-dominated and SF-contaminated AGNs in the normalized \OIII\ velocity dispersion and \OIII\ luminosity plane (Figure \ref{fig:AGN_dominated}). While SF-contaminated sources are mostly radio-weak and extended targets and many of them are potentially not a true radio AGN,
we find no significant difference of the normalized \OIII~velocity dispersion between the two subclasses. The increasing trend of \OIII\ velocity dispersion with increasing \OIII\ luminosity is almost similar except for the highest \OIII~luminosity bin, supporting that in general radio activity may not have an important effect on diving ionized gas outflows.

 %
\begin{figure}[!htb]
\centering
\includegraphics[width=0.95\linewidth]{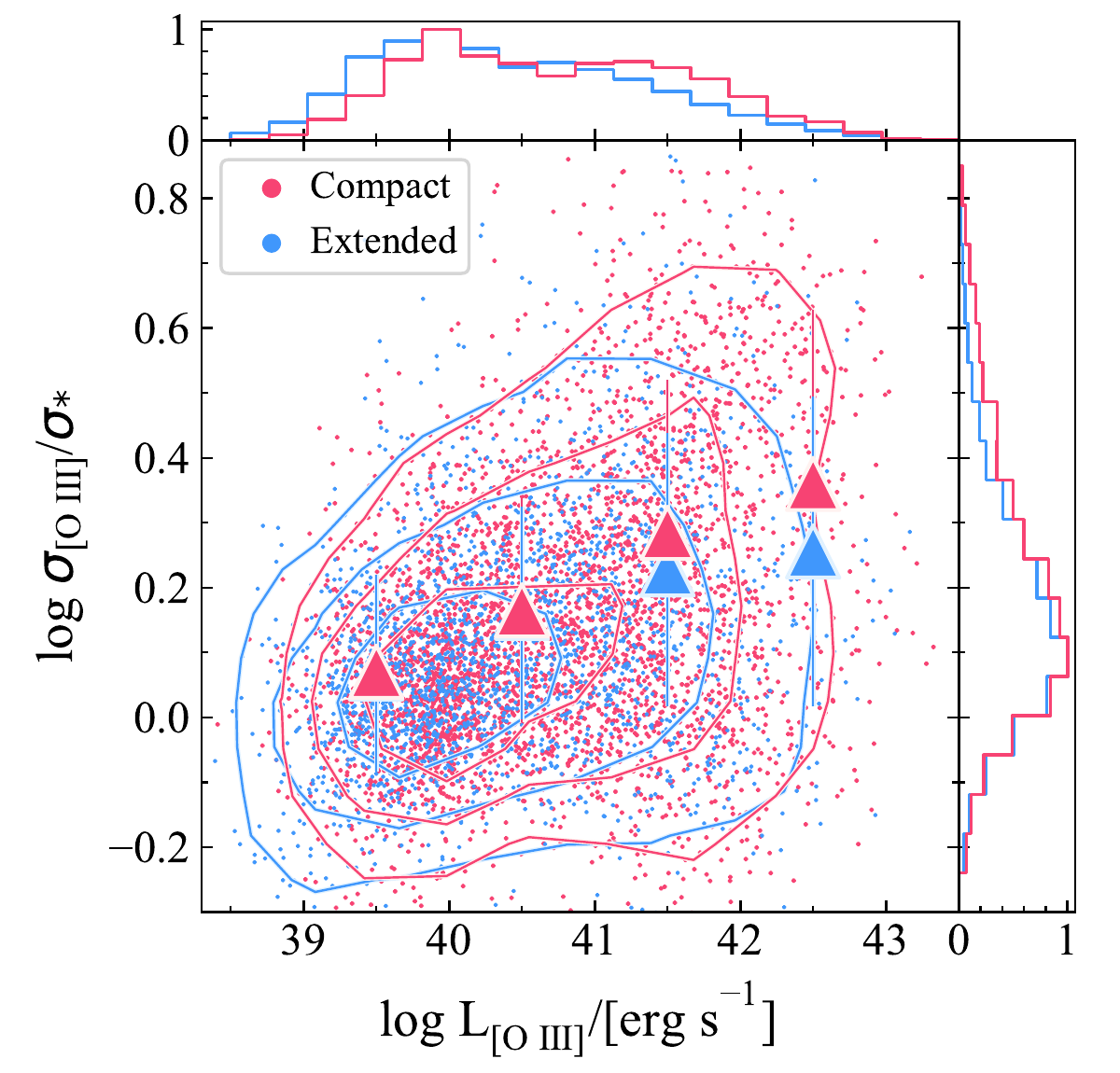}
\caption{Same as in Figure \ref{fig:luminosity}, but for compact (red) and extended (blue) radio sources.}
\label{fig:compactness}
\end{figure}

 %
Third, we investigate the effect of radio morphology by comparing \OIII\ velocity dispersion of compact and extended sources in Figure \ref{fig:compactness}.
The compact radio sources are believed to be young sources and if the age of radio activity is a few $\times$ 10$^{6}$ yr (\citealp{Blundell1999}), this time scale is comparable to the AGN accretion timescale (\citealp{Yu2002}). Thus, it would be interesting to investigate whether compact radio AGNs show stronger outflows than extended sources. We find no significant difference of the mean normalized \OIII~velocity dispersion at a fixed \OIII~luminosity between the two subsamples. Nevertheless, we notice that in the two high \OIII~luminosity bins, compact sources show marginally higher normalized \OIII~velocity dispersion than extended sources. Although the number of targets in these bins is modest, it may indicate that compact jets can play a partial role in driving gas outflows in addition to the accretion-driven outflows only when the accretion luminosity is high.\

Based on the comparison of the \OIII~line width between compact and extended radio sources, \citet{Molyneux2019} argued that young or weak radio jets may be responsible for the broad \OIII~line and extreme gas outflows in compact sources (see also \citealp{Jarvis2019}; \citealp{Santoro2020}). The compact sources in our sample also present slightly broader \OIII~lines than the extended sources over the entire dynamic range of \OIII~luminosities. As Figure \ref{fig:compactness} illustrates, however, once we normalize \dispoiii~ by \dispst\ to correct for the effect of host galaxy gravitational potential, the difference between the compact and extended sources is not evident except for the highest luminosity bins. Note that we find the same results when we use only Type 1 or Type 2 AGNs. These findings indicate no significant evidence that compact sources are directly connected to strong outflows.

 %

\begin{figure}[!htb]
\centering
\includegraphics[width=0.95\linewidth]{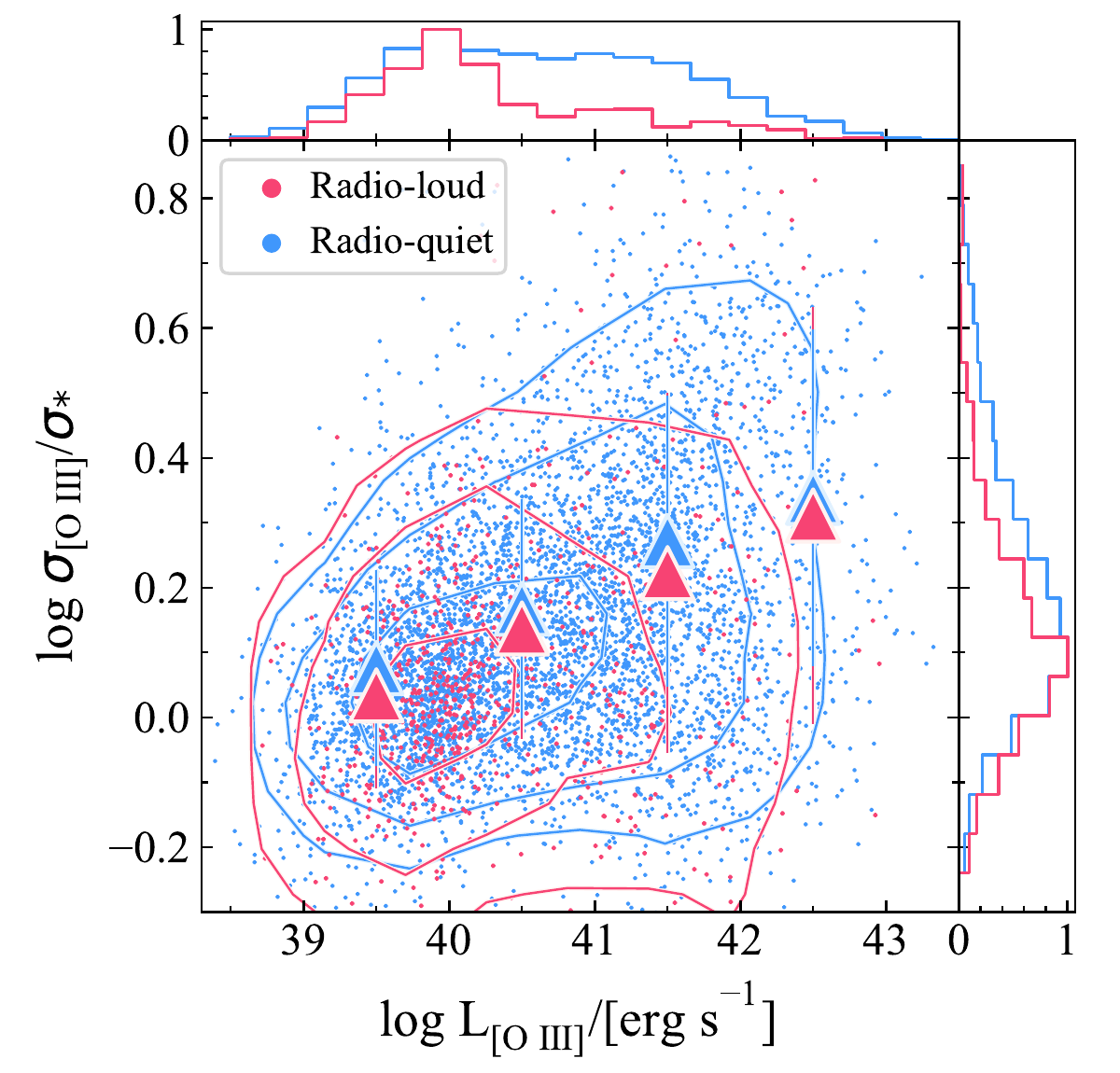}
\caption{Same as in Figure \ref{fig:luminosity}, but for radio-loud (red) and radio-quiet AGNs (blue).}
\label{fig:loudness}
\end{figure}

Last, we investigate whether radio-loudness is connected to gas outflows. 
While RL AGNs preferentially host powerful relativistic radio jets, RQ AGNs are speculated to have weak/no radio jets (\citealp{Urry1995}). Therefore, comparing outflow kinematics between RL and RQ AGNs may provide a clue on the link between outflows and radio jets. In Figure \ref{fig:loudness}, we show the distribution of \dispoiii/\dispst~as a function of \lumoiii~for the RL and RQ AGNs, respectively. It is evident that at any given \lumoiii, there is no significant difference of the mean \dispoiii/\dispst~between RL and RQ AGNs, indicating that RL AGNs show no additional increase of the normalized \OIII~velocity dispersion compared to RQ AGNs. This result support that the radio activity traced by radio-loudness has no direct connection with gas outflows. 

Similar to the compact sources, our RL AGNs exhibit a slight enhancement in \OIII~line broadening. 
Previous studies attribute the broader \OIII~line to the interaction of the ionized gas with the relativistic jets (e.g., \citealp{Berton2016}). However, in comparison with RQ AGNs, host galaxies of RL AGNs host are typically massive elliptical galaxies (e.g., \citealp{McLure2004}; \citealp{Wierzbowska2017}). Thus, the broader \OIII~profile in the RL AGNs is likely due to the strong gravitational potential of their hosts. Our results showed that once the effect of gravitational potential is taken into account, RL AGNs show no enhancement of gas kinematics compared to RQ AGNs.

We test the orientation effect by performing the same analysis for Type 1 and Type 2 AGNs, respectively. 
Since the direction of outflows in Type 1 AGNs is closer to line-of-sight than that of Type 2 AGNs, the inclination angle and the effect of dust extinction in the host galaxy is systematically different between them (see \citealp{Rakshit2018} for more details). Thus, we investigate the difference of \OIII\ velocity dispersion between two radio subclasses using only Type 1 or Type 2 AGNs. We find that the results remain the same, suggesting that the systematic effect of the orientation angle on the outflow kinematics is not significant. \
 
Since the observed radio luminosity can be significantly contributed by SF activity, we repeat the above analysis after correcting for the SF effect. 
We define the corrected radio luminosity (L$_{\rm 1.4\ GHz,\ Corr}$) by subtracting the contribution from SF activity from the observed radio luminosity (i.e., L$_{\rm 1.4\ GHz,\ Corr}$ =  L$_{\rm 1.4\ GHz}$ -  L$_{\rm 1.4\ GHz,\ SF}$). In this process, we exclude 636 sources for which the observed radio luminosity can be entirely attributed to SF (i.e., L$_{\rm 1.4\ GHz,\ Corr} < $ 0). 
We find no significant change, indicating that the systematic uncertainty of AGN radio luminosity due to SF contamination is not significant in exploring the role of radio activity in our analysis.

Based on these results, we conclude that there is no direct link between ionized gas outflows and radio activity traced by\ radio luminosity, compactness, and radio-loudness. While \OIII\ luminosity show a strong correlation with the normalized \OIII\ velocity dispersion, radio activity does not show any significant connection with outflows.

\subsection{Ionized Gas Outflows vs. Radio-Eddington Ratio}\label{sec:radio_edd}

\begin{figure}
\centering
\includegraphics[width=0.99\linewidth]{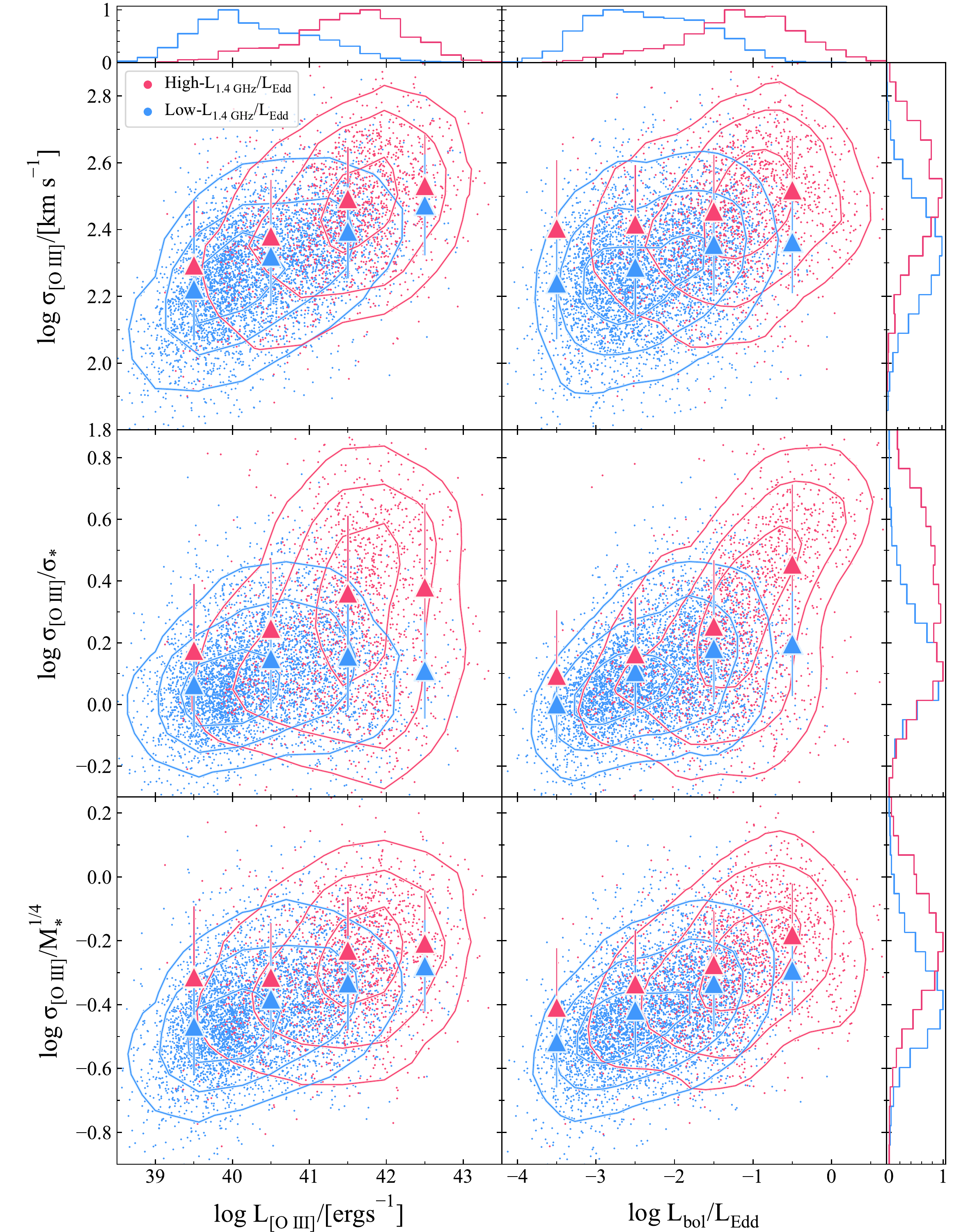}
\caption{The \OIII~velocity dispersion with and without normalization by stellar velocity dispersion and stellar mass, as a function of \OIII~luminosity ($left$) and the bolometric-Eddington ratio (hereafter, Eddington ratio, $right$) for the high radio-Eddington (red) and low radio-Eddington AGNs (blue). The mean value of \OIII~velocity dispersion in each bin is denoted with triangles. The contours enclose 30\%, 70\% and 90\% of targets in each subsample. The error bars indicate 1$\sigma$ dispersion.}
\label{fig:radio_edd}
\end{figure}   

In section, \ref{sec:radio} we find no evidence of a significant role of radio luminosity in driving \OIII~outflows when we divide the sample based on different radio properties. However, in Figure \ref{fig:vvd2}, we detect a clear dependence of \OIII~kinematics on the radio-Eddington ratio. In this section, we further investigate this dependency by dividing the radio AGNs into high and low radio-Eddington ratios. Using the border line at log $\rm L_{1.4\ GHz}/L_{Edd} = -7$, which is the mean value of the total sample, we define high radio-Eddington and low radio-Eddington AGNs.
In Figure \ref{fig:radio_edd} we compare the \OIII~velocity dispersion with \OIII~luminosity (left) and Eddington ratio (right) for these two subsamples. 
On average high radio-Eddington AGNs have higher \OIII\ luminosity and higher Eddington ratio. However, for given \OIII\ luminosity or Eddington ratio, 
high radio-Eddington AGNs exhibit broader \OIII~profile than low radio-Eddington AGNs (top panel). As we discussed in Section \ref{sec:gravity}, this apparent trend may be caused by the gravitational potential of host galaxies. Hence, we normalize \OIII~velocity dispersion with either stellar velocity dispersion (middle) or stellar mass (bottom). Nevertheless, we find that high radio-Eddington AGNs have on average a higher normalized \OIII~velocity dispersion across the entire dynamic range of \OIII~luminosity and Eddington ratio. These results suggest that high radio-Eddington AGNs show stronger gas outflows than low radio-Eddington AGNs, even after correcting for the effect of host galaxy gravitational potential on the \OIII\ line width.
For given \OIII\ luminosity and Eddington ratio, the ionized gas outflows in high radio-Eddington AGNs are on average stronger by a factor of $\sim$ 1.4 and $\sim$ 1.2, respectively, than those in low radio-Eddington AGNs. These results suggest that in addition to the role played by accretion rate, radio activity contributes to driving gas outflows. 

For a consistency check, we repeat the same analysis after excluding SF-contaminated objects, which are a significant fraction of the high radio-Eddington AGNs (i.e., $\sim$ 30\%) and of low radio-Eddington AGNs (i.e., $\sim$ 40\%), finding that the results remain the same. 
Considering the systematic difference of estimating black hole mass and stellar velocity dispersion between type 1 and type 2 AGNs, we also test the trend using only type 1 or type 2 AGNs and find consistent results.

%
   
\subsection{Ionized Gas Outflows vs. AGN Accretion}\label{sec:AGN}

\begin{figure}
\centering
\includegraphics[width=0.99\linewidth]{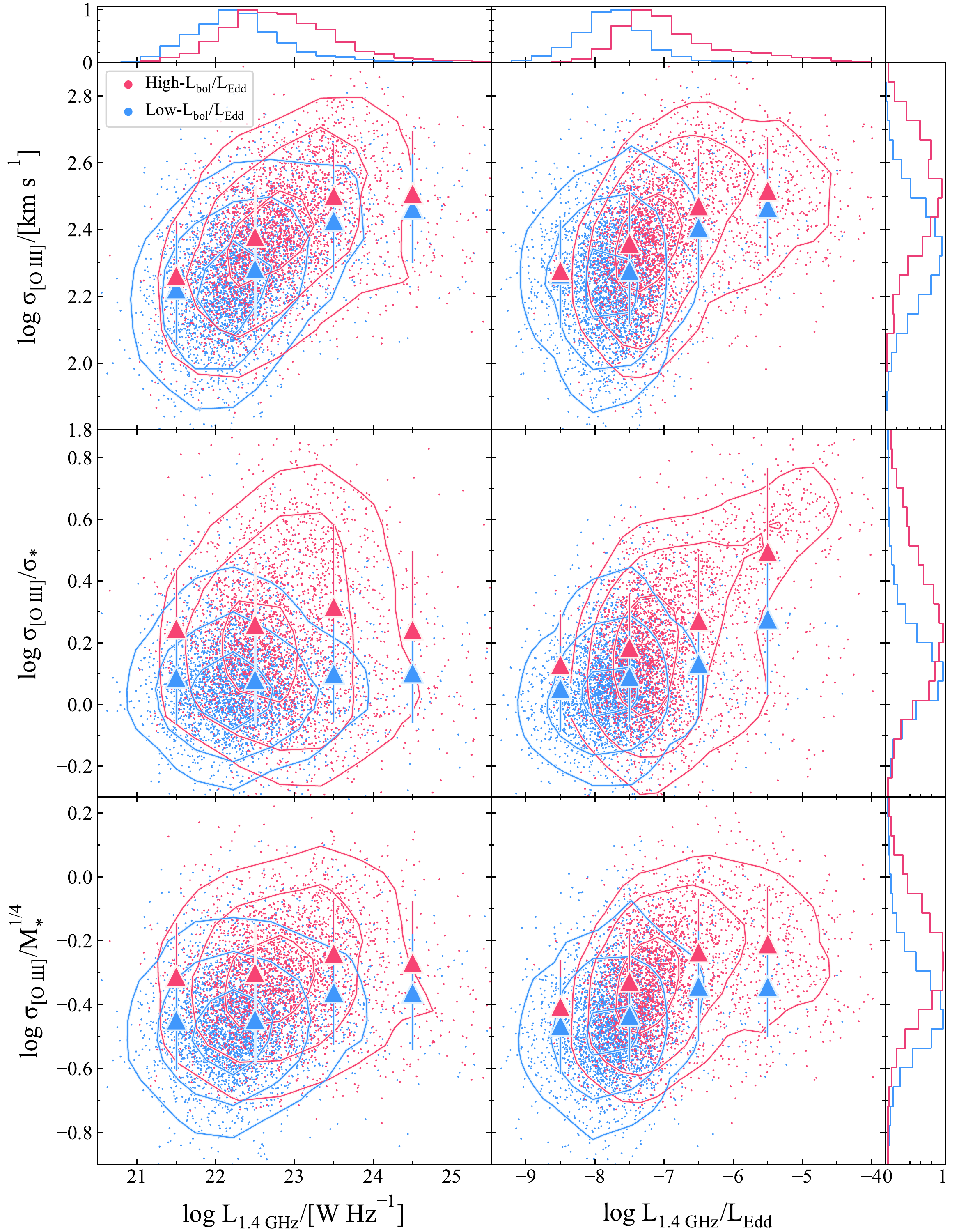}
\caption{The \OIII~velocity dispersion with and without normalization by stellar velocity dispersion and stellar mass, as a function of radio luminosity ($left$) and radio-Eddington ratio ($right$) for the high Eddington (red) and low Eddington AGNs (blue). The mean value of \OIII~velocity dispersion in each bin is denoted with triangles. The contours enclose 30\%, 70\% and 90\% of targets in each subsample. The error bars indicate 1$\sigma$ dispersion.}
\label{fig:optical_class}
\end{figure}   

In this section we investigate the dependency of ionized gas outflows on the accretion luminosity, while we control radio activity. The Eddington ratio of our sample spans $\rm -4.6\ <\ log\  L_{bol}/L_{Edd} <$ 1.1 (see Section \ref{sec:main_par}), and we define 2,848 high Eddington AGNs with $\rm log\ L_{bol}/L_{Edd} > -2$, and 3,185 low Eddington AGNs with $\rm log\ L_{bol}/L_{Edd} < -2$. In Figure \ref{fig:optical_class}, we present the \OIII~velocity dispersion against radio luminosity (left) and radio-Eddington ratio (right) for high and low Eddington AGNs. We find a clear difference of the normalized \OIII~velocity dispersion between the two subclasses at fixed radio emission. For given radio luminosity or radio-Eddington ratio, the ionized gas outflows in high Eddington AGNs are on average stronger by a factor of $\sim$ 1.4 than those in low Eddington AGNs. To check whether this difference depends on AGN type, we use Type 1 and Type 2 AGNs separately, finding that high Eddington AGNs tend to show much higher normalized \OIII~velocity dispersions in both types. A similar conclusion can be reached if we track accretion power by using the \OIII~luminosity. 
Considering the systematic difference of estimating black hole mass and stellar velocity dispersion between type 1 and type 2 AGNs, we repeat the same analysis using only type 1 or type 2 AGNs, finding consistent results. 

Finally, we conduct a simple principal component analysis (PCA) to investigate the importance of features, using the normalized \OIII~velocity dispersion, Eddington ratio, and radio-Eddington ratio as inputs. The results indicate that 80\% of the variance in our dataset can be explained by the first principal component (PC1). According to the absolute values of the eigenvectors' components, Eddington ratio contributes to PC1 slightly more than radio-Eddington ratio, with eigenvector components of 0.72 and 0.66, respectively.

These results suggest that accretion rate traced by Eddington ratio is the main parameter in driving strong gas outflows in radio AGNs and radio activity may have a secondary role in providing an additional effect.

\section{Discussion}\label{sec:discussion}

It has been illustrated that multi-phase outflows could be driven by radio jets by a number of observational studies (e.g., \citealp{Jarvis2019}; \citealp{Husemann2019}; \citealp{Venturi2021}; \citealp{Girdhar2022}; \citealp{Speranza2022}) as well as simulation studies (e.g., \citealp{Wagner2013}; \citealp{Mukherjee2018}; \citealp{Tanner2022}). According to this scenario, a young and compact jet propagates through a clumpy medium, inflating a bubble of shocked gas, creating a multi-phase ISM, and lifting the multi-phase gas outflows, while the final outcome of jet-ISM interaction depends on the jet orientation and jet power (\citealp{Mukherjee2018}). Compared to the radio jets perpendicular to the galaxy disk, the highly inclined jets pass through a larger gas column and interact more strongly with dense ISM. On the other hand, the radiatively driven ionized gas outflows are also reported in the case of radio AGNs based on the spatially-resolved observations (\citealp{Speranza2021}). \
 
Using a large sample of $\sim$ 6000 radio AGNs at $z < 0.4$, we present the kinematics of ionized gas outflows manifested by the \OIII~line profile and investigate the correlation of outflows with radio properties, in order to constrain the role of radio activity in driving gas outflows. We find that accretion rate traced by Eddington ratio plays an essential role in driving ionized gas outflows over the large dynamic range of radio luminosity or radio-Eddington ratio (Section \ref{sec:AGN}),This result presents compelling evidence for a strong connection between accretion rate and gas outflows. 
The radiation emitted by an AGN can strongly interact with the surrounding gas, leading to AGN-driven outflows. Consequently, an increasing trend in the outflow velocity is expected as Eddington ratio increases. Our conclusion is consistent with the previous studies of low-excitation radio galaxies (\citealp{Singha2021}), high-excitation and broad-line radio galaxies (\citealp{Speranza2021}), which show that ionized gas outflows are more likely to be driven by AGN radiation pressure rather than radio jets. 

In contrast, radio properties do not significantly change the correlation between \OIII\ velocity dispersion and \OIII\ luminosity, suggesting that jet-driven outflows, if any, produce much weaker effects on gas kinematics than accretion-induced outflows in radio AGNs.
In particular, we investigate the effect of host galaxy gravitational potential on broadening \OIII\ line width. Since radio luminous AGNs are preferentially hosted by massive galaxies, the gravitational potential of host galaxies is at least partly responsible for broadening emission line profiles observed in spatially-unresolved spectra (e.g., \OIII\ in the SDSS spectra). Therefore, the increasing trend of \OIII~line width with radio luminosity reported in the previous studies \citep[e.g.,][]{Veilleux1991, Mullaney2013, Zakamska2014} may not directly indicate that radio jets are the dominant driver of ionized gas outflows. In fact, we find that the apparent positive correlation between \OIII~velocity dispersion and radio luminosity becomes much weaker (Figure \ref{fig:cum}), when we utilize the relative strength of ionized gas outflows by normalizing \OIII~velocity dispersion with either stellar velocity dispersion or stellar mass, in order to account for the effect of the host galaxy gravitational potential \citep[see also][]{Woo2016, Rakshit2018, Luo2019}. These results indicates the importance of eliminating the effect of the host galaxy gravitational potential in quantifying outflow kinematics.\
 
Nevertheless, we find a significant trend of increasing \OIII\ velocity shift and velocity dispersion with radio-Eddington ratio. At fixed accretion luminosity (traced by \OIII\ luminosity or Eddington ratio) we find a significant different between low and high radio-Eddington AGNs, suggesting that radio activity plays a role in addition to accretion in driving gas outflows. Thus, it is likely that both accretion and radio activity drive gas outflows in radio AGNs, while the relative importance and efficiency of the two mechanisms are necessary to be carefully investigated.

To distinguish the role played by accretion and radio activities, it it necessary to properly compare accretion luminosity and radio power with spatially resolved outflow properties, i.e., outflow size, opening angle, distribution of emission line flux and outflow velocity, etc, for which high signal-to-noise ratio and high spatial resolution data are required. In the near future we plan to investigate accretion-jet connection in driving gas outflows by selecting high radio-Eddington AGNs.

 \

\section{Summary and Conclusions}\label{sec:conclusion}   
In this study, we investigated the connection of ionized gas outflows with radio properties using a large sample of radio AGN at $z < 0.4$.
Our main findings are summarized as follows.\\

$\bullet$
We find that the distribution of the \OIII\ velocity and velocity dispersion increases with increasing radio-Eddington ratio, while the increasing trend is weaker with radio luminosity. These results suggest that radio-Eddington ratio is likely to be a key in investigating the role of radio activity in driving gas outflows.
\\

$\bullet$
We confirm that the positive trend between \OIII~velocity dispersion and radio luminosity becomes much weaker once we normalize \OIII~velocity dispersion by either stellar velocity dispersion or stellar mass, as previously reported by \citealp{Woo2016} and \citealp{Rakshit2018}. These results do not depend on selection of radio samples, emission line fitting methods, and the representation of outflow kinematics (i.e., $\sigma_{\rm OIII}$ or FWHM$_{\rm OIII}$). 
Thus, the effect of host galaxy gravitational potential should be corrected for in order to properly investigate outflow kinematics, particularly for AGNs hosted by massive galaxies.  
\\

$\bullet$
By dividing the radio AGNs into a series of binary subclasses based on four different indicators of radio activity, i.e., radio-luminous/radio-weak, AGN-dominated/SF-contaminated, compact/extended, and radio-loud/quiet, we investigate the effect of radio activity on outflow kinematics. We find that none of these two subclasses exhibit a significant difference of \OIII~kinematics at a given \OIII~luminosity, suggesting that radio-driven outflows may be much weaker than accretion-driven outflows in general.
\\

$\bullet$
Compared to low radio-Eddington ratio AGNs, our findings indicate that high radio-Eddington ratio AGNs have systematically higher \OIII~velocity dispersion over the entire dynamic range of \OIII~luminosity or Eddington ratio, suggesting that radio activity provides additional boost in driving ionized gas outflows in radio AGNs. 
\\

$\bullet$
We find that high Eddington ratio AGNs show systematically higher \OIII~velocity dispersion than low Eddington ratio AGNs at fixed radio luminosity or radio-Eddington ratio. In combination with simple principal component analysis (PCA), we conclude that accretion rate, that is traced by bolometric Eddington ratio plays a major role in driving strong gas outflows.


\acknowledgments
We thank the anonymous referee for detailed comments, which were useful to improve the clarity of the manuscript. This work has been supported by the Basic Science Research Program through the National Research Foundation of the Korean Government (grant No. NRF-2021R1A2C3008486). A.A. acknowledges support from China Scholarship Council for the Ph.D. Program (No. 2017SLJ021244). S.R. acknowledges the partial support of SRG-SERB, DST, New Delhi through grant no. SRG/2021/001334.

\bibliography{bib.bib}
\end{document}